\documentclass[printer]{aa}
\usepackage{graphicx}
\usepackage{epsfig}
\usepackage{subfigure}
\usepackage{amssymb}
\usepackage{txfonts}
\def\mso{\,{\rm M}_\odot}

 \def\kms{\, {\rm km}\, {\rm s}^{-1}}

 \def\simle{\mathrel{\hbox{\rlap{\hbox{\lower4pt\hbox{$\sim$}}}\hbox{$<$}}}}
 \def\simgr{\mathrel{\hbox{\rlap{\hbox{\lower4pt\hbox{$\sim$}}}\hbox{$>$}}}}
 \def\vinf{\, \mathrm{v}_\infty}
 
 \def\vrot{\, \mathrm{v}_{\mathrm{rot}}}
 \def\vcrit{\, \mathrm{v}_{\mathrm{crit}}}
 \def\mdot{\, \dot{M}}
 \def\msoy{\, \mso~{\rm yr}^{-1}}
\begin{document}
   \title{The circumstellar medium around a rapidly rotating, chemically homogeneously evolving
possible gamma-ray burst progenitor}

   \author{A. J. van Marle
          \inst{1,2}
          \and
          N. Langer
          \inst{1}
	  \and
          S.-C. Yoon
          \inst{3}
          \and
          G. Garc{\'i}a-Segura
          \inst{4}
          }

   \offprints{A. J. van Marle}

   \institute{Astronomical Institute, Utrecht University, 
              P.O.Box 80000, 3508 TA, Utrecht, The Netherlands\\
              \email{A.vanMarle, N.Langer@astro.uu.nl}      
         \and
             Bartol Research Institute, University of Delaware, 
             102 Sharp Laboratory, Newark, 19716 DE, Delaware, USA\\     
             \email{marle@udel.edu}               
         \and
              Astronomical Institute "Anton Pannekoek", University of Amsterdam, 
              Kruislaan 403, 1098 SJ, Amsterdam, the Netherlands\\
              \email{scyoon@science.uva.nl}
         \and
             Instituto de Astronom{\a'i}a-UNAM, 
             APDO Postal 877, Ensenada, 22800 Baja California, Mexico\\
             \email{ggs@astrosen.unam.mx}
             }

   \date{Received <date> / Accepted <date>}

   \abstract{Rapidly rotating, chemically homogeneously evolving massive stars are considered to be progenitors of 
long 
gamma-ray bursts.}
   {We present numerical simulations of the evolution of the circumstellar medium around a rapidly rotating 
20~$\mso$ star at a metallicity of $Z=0.001$. Its rotation is fast enough to produce  
quasi-chemically homogeneous evolution. While conventionally, a star of 20~$\mso$ would not evolve into
a Wolf-Rayet stage, the considered model evolves from the main sequence directly to the helium main sequence.}
   {We use the time-dependent wind parameters, such as mass loss rate, wind velocity and rotation-induced
wind anisotropy from the evolution model 
as input for a 2D hydrodynamical simulation.} 
   {While the outer edge of the pressure-driven circumstellar bubble is spherical, the 
circumstellar medium close to the star shows strong non-spherical features during and after the periods of 
near-critical rotation.}
   {We conclude that the circumstellar medium around rapidly rotating massive stars differs considerably from 
the surrounding material of non-rotating stars of similar mass. 
Multiple blue-shifted high velocity absorption components 
in gamma-ray burst afterglow spectra are predicted. 
As a consequence of near critical rotation and short
stellar evolution time scales during the last few thousand years of the star's life, 
we find a strong deviation of the circumstellar density profile in the polar direction from the $1/R^2$ 
density profile normally associated with stellar winds close to the star.}
	     
   
  \titlerunning{Circumstellar medium around a chemically homogeneous star}
  \authorrunning{van Marle et al.}
  \maketitle
%

\section{Introduction}
Within the collapsar picture (Woosley~\cite{W93}),
long gamma-ray bursts (GRB) are thought to be produced at the end of the evolution of
massive stars, if their rapidly rotating iron cores collapse to form black holes. 
The circumstellar medium (CSM) around GRB progenitors is interesting because 
it determines the GRB afterglow evolution (Chevalier \& Li \cite{CL00}, Panaitesku \& Kumar \cite{PK01, PK02}, 
Chevalier et al. \cite{CLF04}, Ramirez-Ruiz et al. \cite{RGSP05}, Nakar \& Granot \cite{NG06}), and because it may 
produce distinct 
absorption features in GRB afterglow spectra (Mirabal et al. \cite{Metal03}, Schaefer et al. \cite{Setal03}, Fiore 
et al. \cite{Fietal05}, Starling et 
al. \cite{Setal05}). In previous papers, van Marle et al. (\cite{MLG05b}, \cite{MLAG06}, \cite{MLG07}) have 
computed the CSM 
evolution of various types of supernova and GRB progenitors 
without including the effect of rotation of the central star. 

It has been recently suggested that rapidly rotating, chemically homogeneous massive stars
evolve into long GRBs (Yoon \& Langer \cite{YL05}; Woosley \& Heger \cite{WH06}; Yoon et al. \cite{YLN06}; 
Cantiello et al. \cite{CYLL07}). 
This type of evolution is thought to occur at low metallicity, where stellar winds are weaker and angular momentum 
loss by winds (Langer~\cite{L98}) is therefore limited. Since models of chemically homogeneously evolving stars 
avoid a redward evolution in the HR diagram, but rather evolve from the main sequence directly to the helium main 
sequence, the structure of their circumstellar medium is expected to
be different from that of Galactic Wolf-Rayet stars, which are thought to be the descendants of red supergiants or 
Luminous Blue Variables (Garcia-Segura et al. \cite{GML96, GLM96}).
Additionally, the chemically homogeneous stellar models remain very rapid rotators throughout their
lives, and encounter various stages of near-critical rotation (Yoon \& Langer \cite{YL05}).
Wind anisotropies may therefore play an important role in shaping the CSM of these stars.

Here, we undertake an effort to understand the evolution of the circumstellar medium around such stars.
We perform detailed numerical simulations of its time evolution, were we use a detailed stellar
evolution model as input. We employ a prescription of the anisotropy of the mass outflow and wind velocity
as a function of the rotation of the central star, and adopt this, together with the time dependent
global wind properties and ionizing photon fluxes, as the inner boundary condition for our
CSM evolution model. As result, we obtain the density, velocity and temperature structure of the CSM
of our stellar model, from the main sequence to the pre-SN stage. 
   
   In Sect.~\ref{sec-evol} we describe the evolution of the properties of the chemically-homogeneously
   evolving stellar model which is used as input for our hydrodynamic calculations. 
   In Sect.~\ref{sec-num} we discuss the numerical method used to simulate the hydrodynamical 
evolution of the CSM.
   The results of our simulation are presented in Sect.~\ref{sec-result1}.
   In Sect.~\ref{sec-spectrum}, we show the absorption profiles that would result from the morphology of the 
circumstellar medium when the star explodes. Section~\ref{sec-dprof} deals with the density profile of the 
circumstellar medium, which is important for the lightcurve of a GRB afterglow.
   In Sect.~\ref{sec-concl} we discuss our results.

   \begin{figure}
   \centering
   \resizebox{\hsize}{!}{\includegraphics[angle=-90]{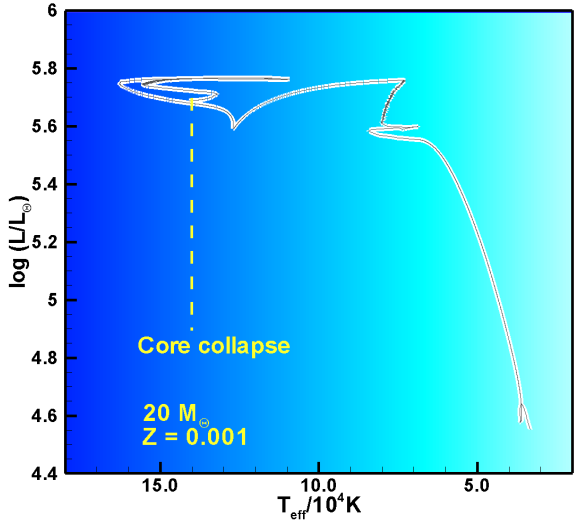}}
      \caption{Evolution the rapidly rotating 20 $\mso$ model at $Z=0.001$ which is used as
    input for the CSM simulations in the Hertzsprung-Russel diagram. The model is similar to the ones used in Yoon 
et al. (\cite{YLN06}). Its initial rotational velocity is 414~km/s, which is about 60\% of the critical velocity. 
The track starts at 
the zero age main sequence and ends at the pre-supernova stage. 
     }
     \label{fig:HRdiag}
   \end{figure}

   \begin{figure}
   \centering
   \resizebox{\hsize}{!}{\includegraphics[angle=-90]{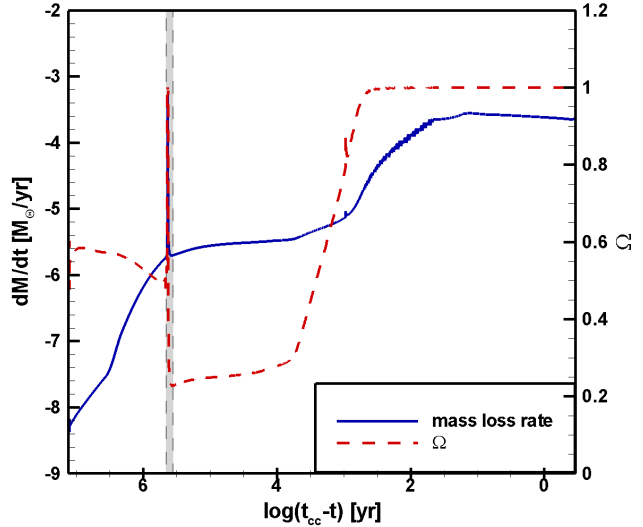}}
	\caption{Latitudinally averaged mass loss rate, and $\Omega$ ($=\vrot/\vcrit$) for our 20 $\mso$ sequence,
    as a function of time.
   The horizontal axis shows the logarithm of the time before core collapse ($t_{cc}$).
        The gray area highlights the time interval which is explored in Figs.~\ref{fig:2d1} to \ref{fig:2d3} 
below.
	}
   \label{fig:mloss1}
   \end{figure}

   \begin{figure}
   \centering
   \resizebox{\hsize}{!}{\includegraphics[angle=-90]{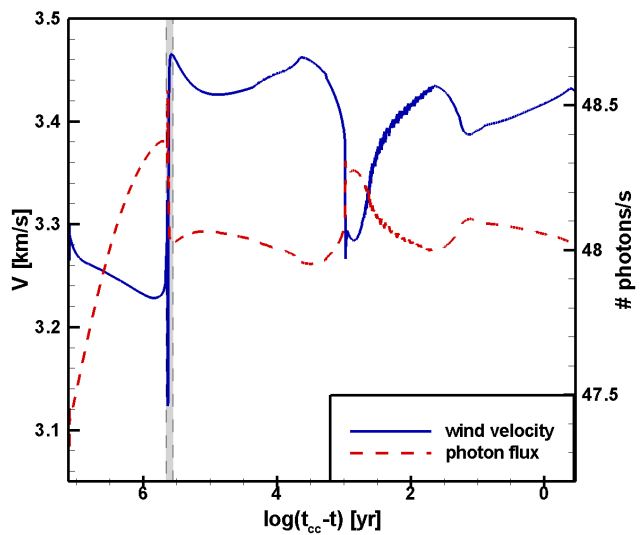}}
	\caption{Number of high energy photons, and terminal wind velocity of our 20 $\mso$ sequence star,
    as a function of time. The horizontal axis shows the logarithm of the time before core collapse ($t_{cc}$).
 The gray area highlights the time interval explored in Figs.~\ref{fig:2d1} to \ref{fig:2d3} below.
	}
   \label{fig:mloss2}
   \end{figure}

   \begin{figure}
   \centering
   \resizebox{\hsize}{!}{\includegraphics[angle=-90]{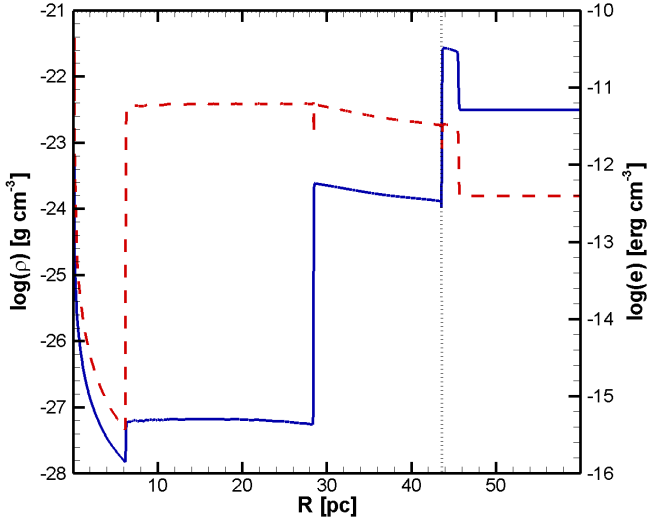}}
      \caption{The result of a 1D hydrodynamic simulation of the CSM around our 20 $\mso$ star, 
 just before the star reaches critical 
rotation for the first time ($4.644\times 10^5$yr before core collapse; cf. Fig.~2).
      The solid line shows the mass density, the dashed line the internal energy density.
      Moving outward from the star, we first encounter the free streaming wind, then the 
shocked wind material ($6~\lesssim~R~\lesssim~30$~pc), the photo-ionized ISM 
($30~\lesssim~R~\lesssim~43$~pc), the shell of shocked ISM ($R~\simeq~45$~pc) and finally, the 
unperturbed ISM. 
      The dotted line marks the Str{\a"o}mgren radius. 
      We use this point in time to start the 2D simulation.
      }
     \label{fig:ms}
  \end{figure}

\section{The stellar evolution model}   
   \label{sec-evol}
   If a star rotates very rapidly, chemical mixing may occur on such a short timescale that no 
significant chemical gradient can be established (Maeder \cite{M87}). 
   Since new material is mixed into the core at all times, core hydrogen burning in these 
stars will continue until all hydrogen in the star is exhausted. 
   Due to this mixing, the star becomes a Wolf-Rayet star during the core-hydrogen burning phase and never swells 
up to become a red supergiant.
   At core hydrogen and core helium exhaustion, the star contracts to reach the 
necessary temperature for the next burning stage. 
   As the star shrinks, its rotation rate increases due to angular momentum conservation. 
   Once the star reaches critical rotation, it can not spin-up any further. 
   Instead, it loses mass at a highly increased rate. 
   With the mass it also loses angular momentum, until the spin velocity drops below the 
critical rotation velocity.
   
   At solar metallicity, massive stars have such a high mass loss rate that they  
lose large amounts of angular momentum (Langer \cite{L98}).
   As the star is spun down, the mixing time increases such that a chemical gradient can be 
established between the core and the envelope.
   However, at low metallicity, the mass loss rate even of Wolf-Rayet stars is considerably lower 
(Nugis \& Lamers \cite{NL00}; Crowther et al. \cite{CDHAF02}; Vink \& de Koter \cite{VK05}),
and it is possible that the star remains quasi-chemically homogeneous throughout core hydrogen burning
(Yoon \& Langer \cite{YL05}; Woosley \& Heger \cite{WH06}).

   As an input for our hydrodynamic CSM study, we use the results of a stellar evolution calculation
for a 20$\,\mso$ star with a metallicity of $Z=0.001$ (cf. Yoon et al. \cite{YLN06}). The initial
equatorial rotational velocity of the model is  414~km/s. The input physics in the stellar evolution
calculation are as in Yoon et al. (\cite{YLN06}), but without stellar wind enhancement due to
increased CNO surface abundances. 
Fig.~\ref{fig:HRdiag} shows the evolutionary track of our star in the 
Hertzsprung-Russel diagram; as a result of the quasi-chemically homogeneous evolution,
the star evolves from the main sequence directly into the Wolf-Rayet regime.

   The mass loss history of chemically homogeneously evolving stars is different from that 
of non-rotating or slowly rotating stars. 
   The latter have a low mass loss rate during their main sequence, and a high 
mass loss rate during the red supergiant stage. 
   The mass loss history of our homogeneous star is shown in Fig.~\ref{fig:mloss1}. 
   Mass loss is low, except during two short phases. 
   During these phases the star reaches critical rotation (dashed line in Fig. \ref{fig:mloss1}).
   This happens for the first time during the overall contraction of the star 
following core hydrogen exhaustion. During this stage, the star loses about ${2 \mso}$ 
in an equatorial outflow, in order to avoid overcritical rotation. 
   Upon core helium exhaustion, the star shrinks again toward carbon ignition,
and approaches critical rotation again.
   However, this time the star reaches the end of its evolution and explodes as a supernova 
before its rotational velocity can drop again. The amount of mass lost in the second phase of
near-critical rotation is ${0.05 \mso}$.

   The terminal wind velocity used for our hydrodynamical calculations does not follow directly from 
the stellar evolution model. 
   We assume here that the wind is radiation driven, which leads us to the following formula
   \begin{equation}
      \vinf = \sqrt{\frac{2 \beta G M}{R}} \sqrt{1-\Gamma_\mathrm{E}}
   \end{equation}
   e.g. Cherchneff \& Tiellens (\cite{CT94}), where $\Gamma_\mathrm{E}=L/L_\mathrm{E}$, the 
ratio between stellar luminosity and Eddington luminosity and 
   $\beta$ is a parameter related to the surface temperature of the star (Eldridge et al. 
\cite{EGDM05}).
   The resulting wind velocity is shown in Fig.~\ref{fig:mloss2}. Due to the compactness
of the stellar model throughout its evolution, the wind speed is at all times of the order
of a few thousand kilometers per second. However, pronounced minima occur at the times
of near critical rotation. 
   
   The values for the mass loss rate and wind velocity shown in Figs.~\ref{fig:mloss1} and 
\ref{fig:mloss2} are spatial averages. 
   Rapidly rotating stars have winds that depend strongly on the latitude.
   As long as the rotational velocity is well below critical, this effect is not important. 
   However, as the rotational velocity increases, so does the asphericity of the wind. 
   We use the equations found by Bjorkman \& Cassinelli (\cite{BC93}, from here 
BC) for winds from rapidly rotating stars (see Sect.~3). 
   This model assumes that for a rapidly rotating star, the wind will be denser at the equator 
than at the poles, while the wind velocity will be higher at the poles than at the equator
(cf. Sects.~\ref{sec-num} and \ref{sec-result1}).

   Since we include the effects of photo-ionization on the CSM, we have to find
the number of high energy photons emitted by the star.
   We approximate this by treating the star as a black body radiator.
   This means that the number of photons emitted per surface area with a certain wavelength only depends on
the surface temperature of the star, according to Planck's formula.
   We simply integrate the total number of photons above the hydrogen ionization threshold
(13.60~eV) and correct this number for the diminishing photo ionization cross section at
higher energies as given by Osterbrock (\cite{O74}).
   The resulting number of ionizing photons radiated by the star are shown in Fig.~\ref{fig:mloss2}.

  \begin{figure*}[H]
   \centering
   \mbox{\subfigure{\epsfig{figure=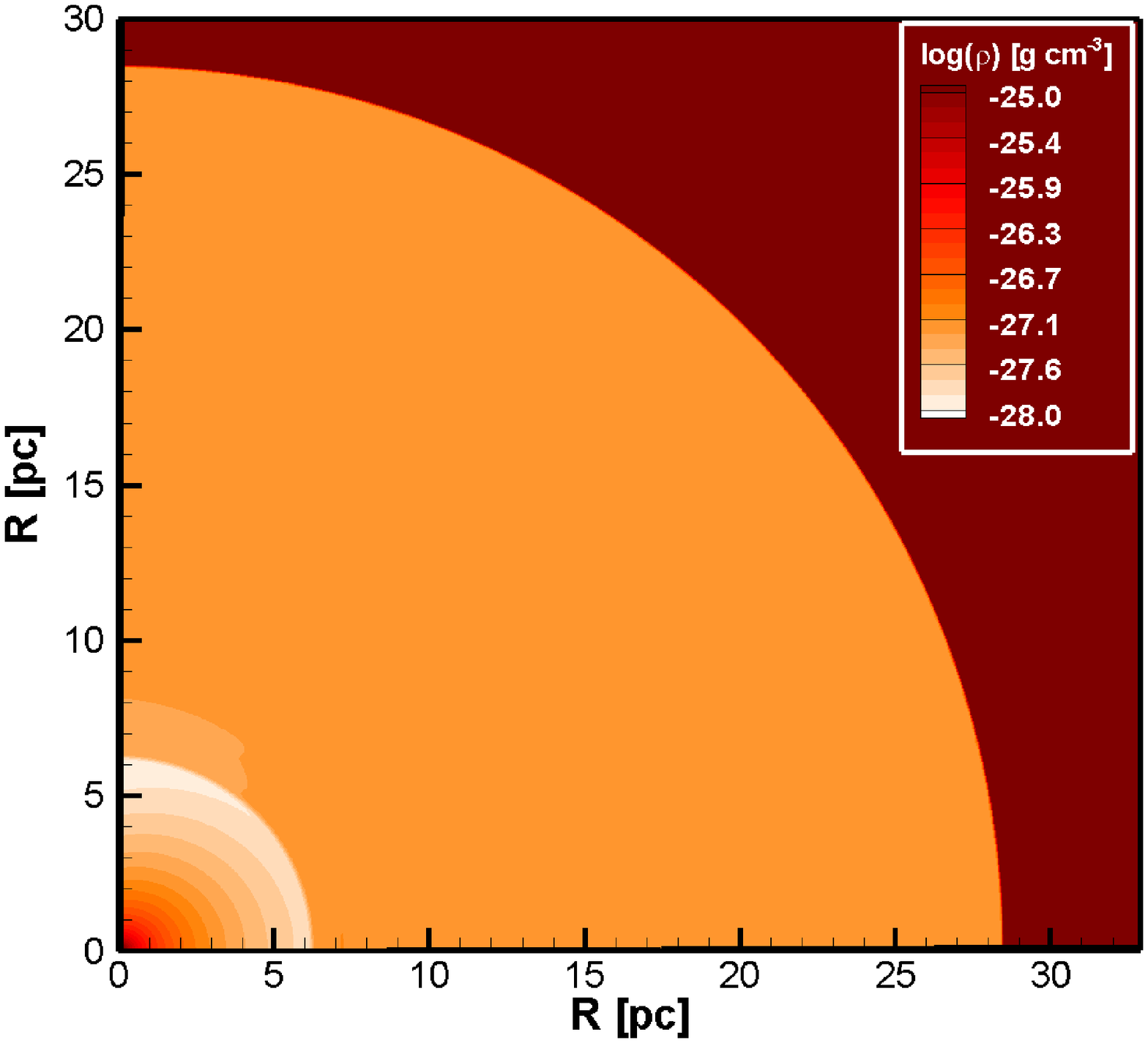,angle=-90,width=0.50\textwidth}}\quad
         \subfigure{\epsfig{figure=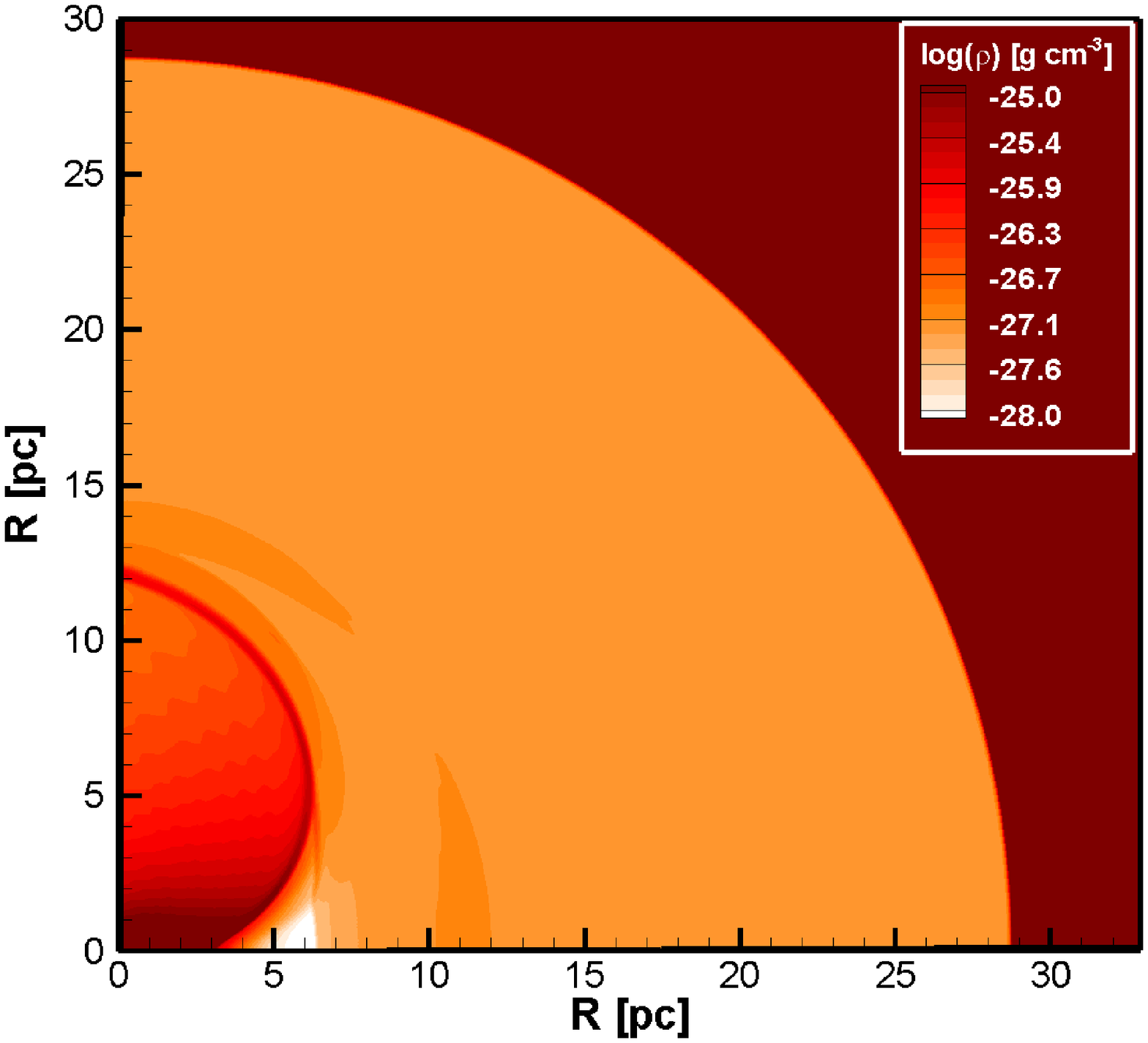,angle=-90,width=0.50\textwidth}}
	 }
      \caption{Density of the CSM at the beginning of the 2D simulation ($4.565\times10^5$yr before core 
collapse). 
      The morphology of the CSM on the left is still similar to that shown in Fig. 
\ref{fig:ms}. 
      However, the wind termination shock has become slightly aspherical due to the angular dependence 
of the ram pressure of the wind. 
      The  Str{\a"o}mgren radius lies outside these figures on the inside of the main 
sequence shell ($R~\simeq~45$~pc).
      In the second frame ($4.327\times10^5$yr before core collaspe), the star has been at critical rotation 
for some time. 
      Close to the star the mass of the wind is concentrated around the equator. 
      However, the wind termination shock reaches its maximum radius at the poles, since the 
wind velocity reaches its maximum in that direction. 
      The increased mass loss rate has created a shell at the inner boundary of the hot 
bubble. 
     }
     \label{fig:2d1}
   \end{figure*}

\section{Numerical method}
\label{sec-num}
We use the ZEUS 3D code (Stone \& Norman \cite{sn92}, Clark \cite{c96})  for our simulations.
  Our numerical method is the same as used by Garc{\a'i}a-Segura et al. (\cite{GML96} \& 
\cite{GLM96}) and in our previous papers (van Marle et al. \cite{MLG05a}, \cite{MLG05b} and \cite{MLG07}). 
  This means that we start the simulation in one dimension and map the result onto a two 
dimensional grid when it becomes necessary.
  Even in the case of a rotating star it is possible to do this.
  Although the wind is aspherical, the main sequence shell is driven by the thermal pressure 
of the shocked wind material (Weaver et al. \cite{WCMSM77}). 
  Therefore, the outer shell of the main sequence bubble will still be spherical, rather than 
ellipsoid. 
  The wind termination shock of course is not spherically symmetric, since the ram pressure of 
the wind depends on the polar angle.
  However, this is no problem, if we start the 2D simulations in time for the gas to 
compensate for this change.
  
  In the 2D simulations we use the same equations for the angle dependence of the wind 
parameters at large distances from the star as were used by Langer et al. (\cite{LGM99}) and  Chi\c{t}\v{a} et al. 
(\cite{CMLG07}). 
  These are based on the model in BC.
  So the wind velocity is:
  \begin{equation}
  \label{eq:vel}
      \vinf(\theta) =  \sqrt{\frac{2 A \beta G M (1-\Gamma_\mathrm{E})}{R}}\, (1-\Omega 
\sin{\theta})^\gamma,
  \end{equation}
  while the wind density is:
  \begin{equation}
  \label{eq:dens}
      \rho(\theta) = \frac{(\alpha/2) B \mdot  (1-\Omega \sin{\theta})^\xi}{4\pi R^2 
\vinf(\theta)},
  \end{equation}
  where we set the parameters defined in BC to $\gamma~=~0.35$ and $\xi~=~-0.43$. 
  The effect of the rotational velocity is included in the parameter $\Omega~=~\vrot/\vcrit$.
  Finally, the parameter $\alpha$ is defined by:
  \begin{equation}
      \alpha=\biggl[\cos{\phi'} + \cot^2\theta \, \biggl(1+\gamma\frac{\Omega 
\sin{\theta}}{1-\Omega \sin{\theta}}\biggr)\, \phi' \sin{\phi'}\biggr]^{-1},
  \end{equation}
  with $\phi'~=~\Omega \sin{\theta} \vcrit/[2\sqrt{2}\, \vinf(\theta)]$.
  Note that for $\Omega \geq 0.995$ these equations do not have a valid solution, since 
$\phi'$ becomes larger than $\pi/2$, which would start to reverse the process of concentrating 
the mass loss toward the equator.
  Therefore, we set $\Omega \leq 0.99$ in our simulations.
  The correction factors $A$ and $B$ serve to make sure that the average stellar mass loss and 
wind velocity remain the same as in the 1D approximation, by averaging the latitude dependent 
parts of Eqs. \ref{eq:vel} and \ref{eq:dens} over the surface area of the star and normalizing 
this to 1. 
  They are calculated by integrating Eqs.~\ref{eq:vel} and \ref{eq:dens} over the surface of 
the star and normalizing the result to the average values shown in Fig. \ref{fig:mloss1} and 
\ref{fig:mloss2}.
 For a different approach to the problem of winds from rotating stars and the effect on the CSM see Eldridge 
(\cite{E07}).
  
  In our simulations we include the effect of photo-ionization in the same way as van Marle et al.(\cite{MLG05b, 
MLG07}).
  This means that we calculate the Str{\a"o}mgren radius along each radial grid line. 
  All matter within this radius is presumed to be fully ionized, while all matter outside the 
Str{\a"o}mgren radius is considered neutral.
  In a completely realistic model, the photon count would depend on the latitude, since the 
star is no longer spherical at high rotational velocity. 
  However, since our treatment of the photo-ionization is only an approximation, 
we treat the photon field emitted by the star as if it were spherically symmetric.
  A more sophisticated way of treating the photo-ionization by massive stars was presented by 
Arthur \& Hoare (\cite{AH06}) and Mellema et al. (\cite{MAHIS06}).
  
  The star has two periods in which the wind is strongly aspherical. 
  The first lasts about 10\,000 years and the mass shed during this phase will have enough time to spread 
throughout the 
entie circumstellar bubble.
  However, the second period is very short and is immediately followed by the supernova 
explosion. 
  Therefore, its effects are only visible close the star. 
  This means that we have to perform a special simulation of the CSM close to the star for the 
last few thousand years of its evolution. 
  Fortunately, due to the short period followed immediately by core collapse, the effects of this second period 
can only be 
felt close to the star. 
  Therefore, this new simulation only has to cover the free-streaming wind region.  
  The hot bubble of shocked wind material, which is typical of wind blown bubbles (Weaver et 
al. \cite{WCMSM77}) does not have to be taken into account.
  This makes the calculation easier since a simple outflow boundary can be used at maximum 
radius rather than the more complicated construction used by Garc{\a'i}a-Segura et al 
(\cite{GML96, GLM96}), for a situation where the outer boundary was inside the hot 
bubble.

   \begin{figure*}  
   \centering
   \mbox{\subfigure{\epsfig{figure=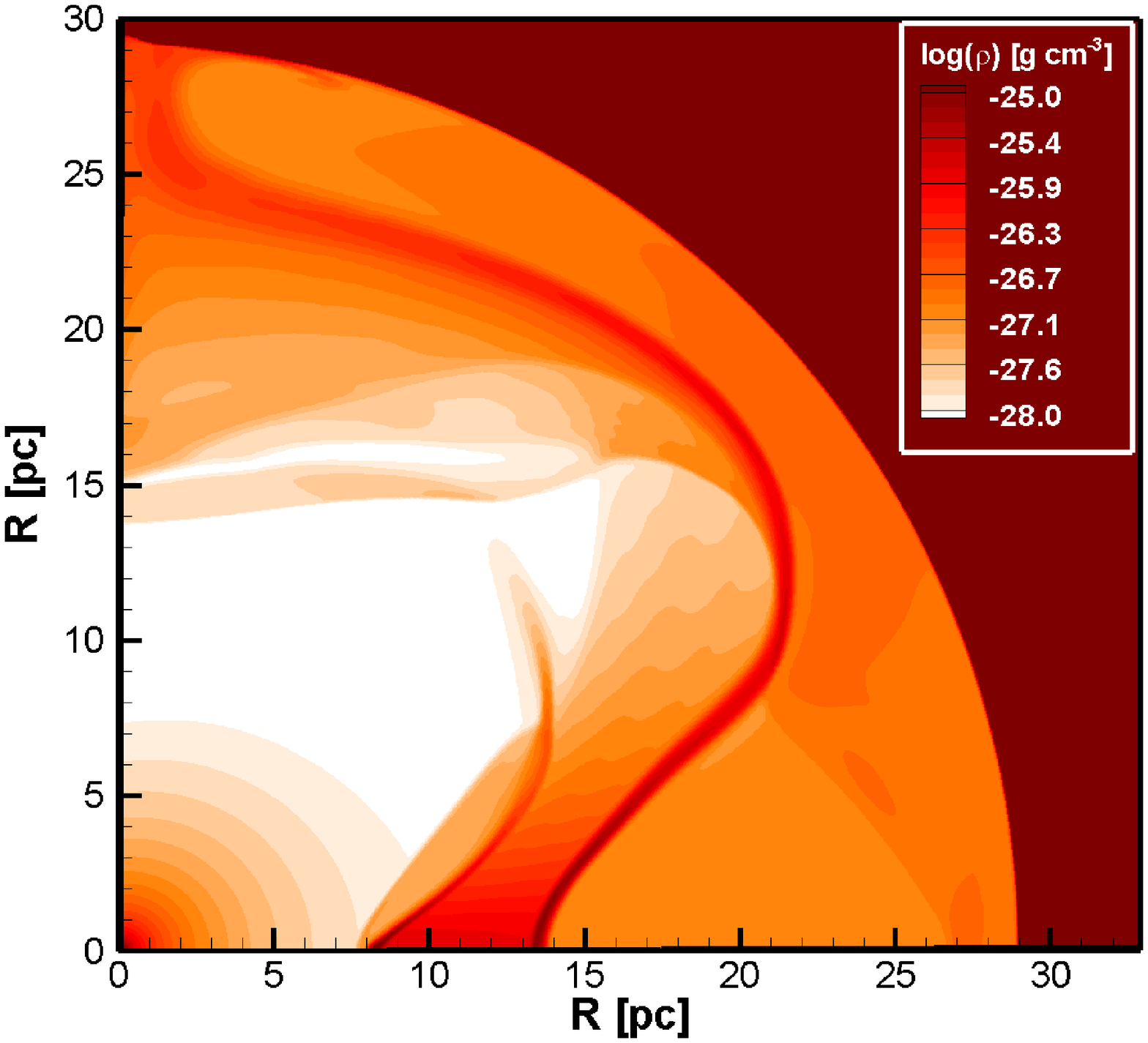,angle=-90,width=0.50\textwidth}}\quad
         \subfigure{\epsfig{figure=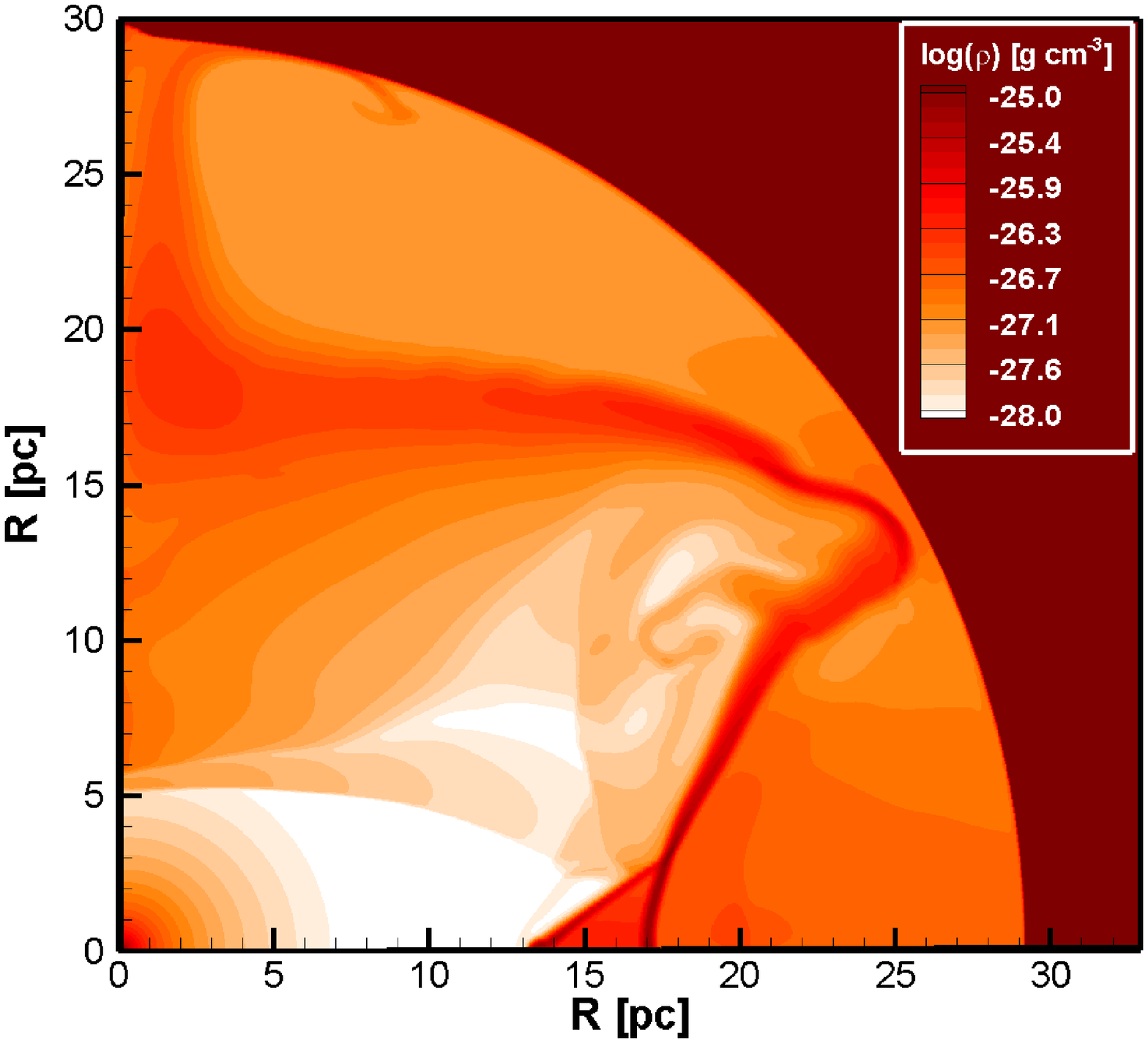,angle=-90,width=0.50\textwidth}}
          }
      \caption{Similar to Fig. \ref{fig:2d1}, after the period of critical rotation has 
passed. 
     On the left ($4.169\times10^5$yr before core collaspe), the star no longer rotates at critical velocity 
and the wind is closer to 
spherical symmetry. 
     At this time the wind, which has become much faster, is in the process of sweeping up the 
surrounding matter. 
     This is easiest toward the poles, since the disk around the equator contains most of the 
mass and is more difficult to move.
     Note the double shell at the outer edge of the free-streaming wind. 
     The outer shell ($R~\simeq~15$~pc at the equator) was caused by increased mass loss and 
lower wind velocity during the period of critical rotation. 
     The inner shell ($R~\simeq~8$~pc at the equator) is the result of the later, faster wind 
sweeping up the disk.
     Eventually, the wind breaks out of the shell, which was formed at the wind termination 
shock. 
     The latitude at which this happens depends on the mass and velocity distribution, both 
during and after the period of critical rotation. 
     The disk-like structure around the equator deflects the wind of the next phase upward, 
causing increased ram pressure at higher latitude. 
    Eventually the disk itself is swept up by the wind as can be seen on the right ($4.089\times10^5$yr before 
core collapse). 
    The pressure equilibrium between the ram pressure of the wind and the thermal pressure of the shocked wind 
bubble changes rapidly during this period. This causes the wind termination shock to move quickly.}
     \label{fig:2d2}
   \end{figure*}

   \begin{figure*}
   \centering
   \mbox{\subfigure{\epsfig{figure=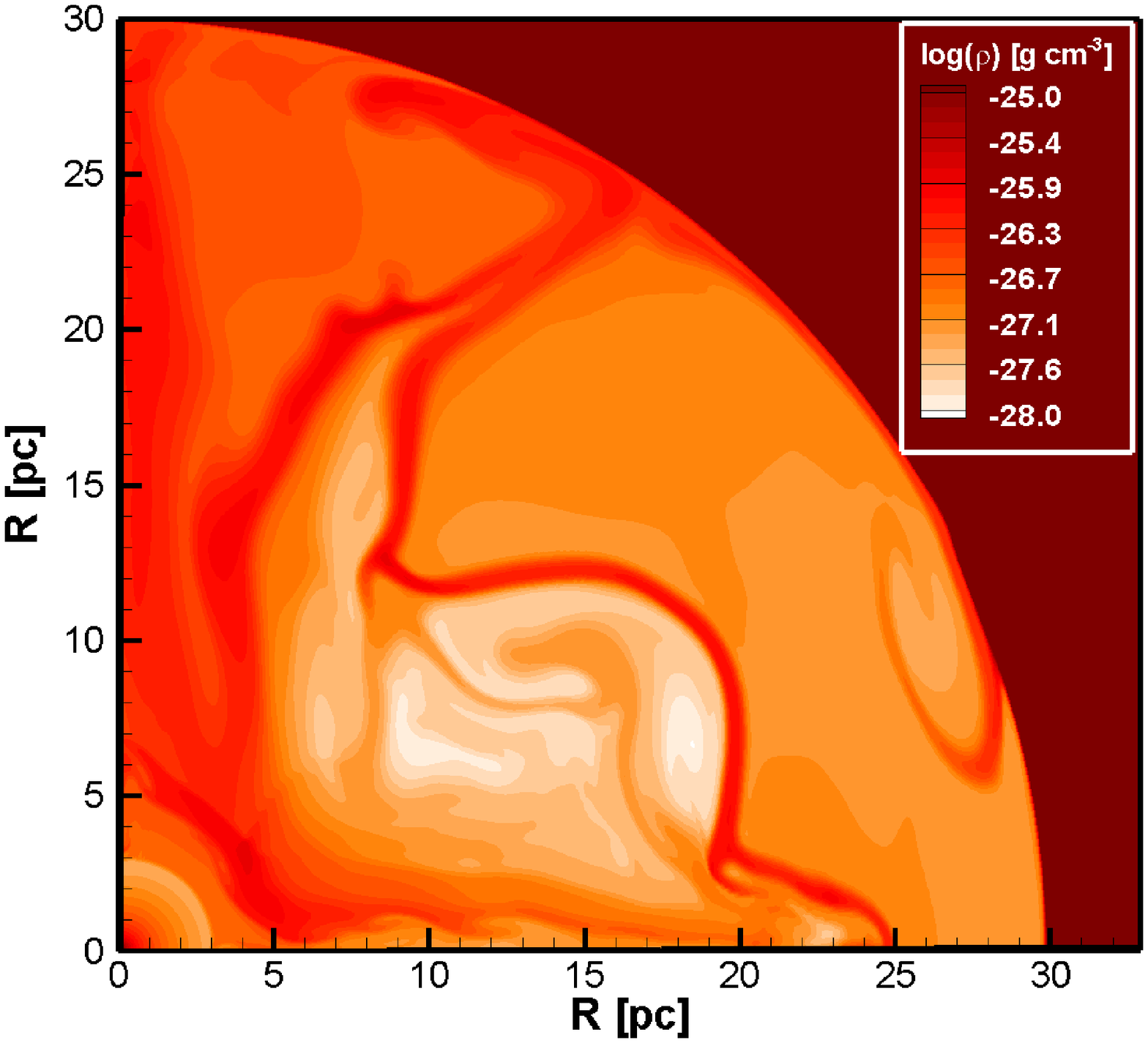,angle=-90,width=0.50\textwidth}}\quad
         \subfigure{\epsfig{figure=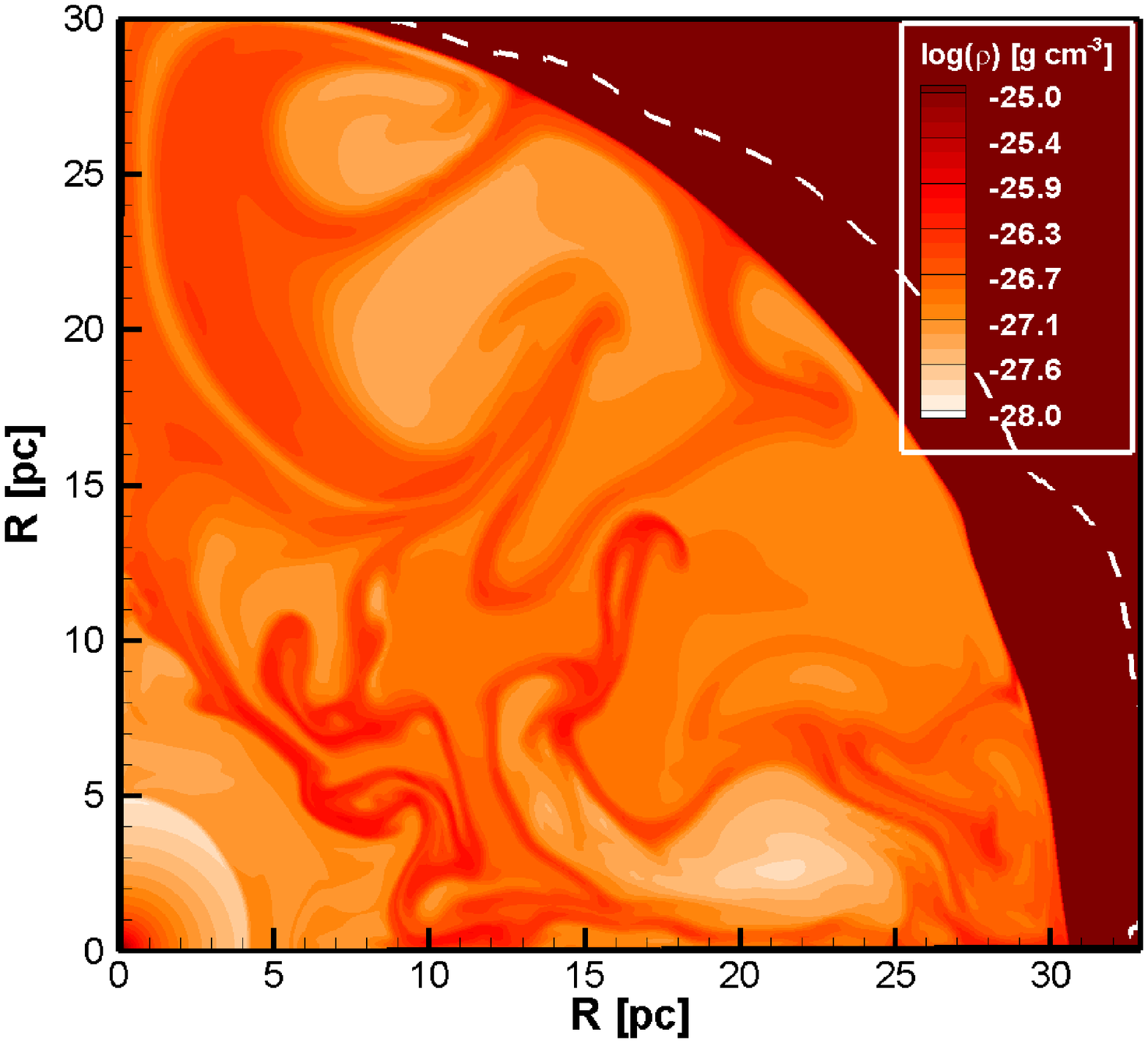,angle=-90,width=0.50\textwidth}}
          }
      \caption{Similar to Figs. \ref{fig:2d1} and \ref{fig:2d2}. 
     The shell has been completely fragmented and a new wind termination shock is being established between the 
free streaming wind and the shocked wind material ($simeq$~3~pc in the figure on the left, $3.852\times10^5$yr 
before core collapse).
    The shocked wind material is highly turbulent.      
     In the right hand frame ($3.614\times10^5$yr before core collapse), a new equilibrium has been reached 
between the thermal pressure 
in the hot bubble and the ram pressure of the wind. 
     The fragments of the shell dissipate into the surrounding material and the turbulent 
movement in the bubble starts to disappear. 
     By the time the star once again reaches critical rotation, the effects of the first 
critical rotation episode will be gone. 
     Because the amount of material that has to be photo-ionized has increased during the period of critical 
rotation as well as because of the higher density loss rate of the wind, the  Str{\a"o}mgren radius (dashed 
white line) has nearly receded to the edge of the shocked wind material.}
     \label{fig:2d3}
   \end{figure*}

   \begin{figure}
   \centering
   \resizebox{\hsize}{!}{\includegraphics[angle=-90]{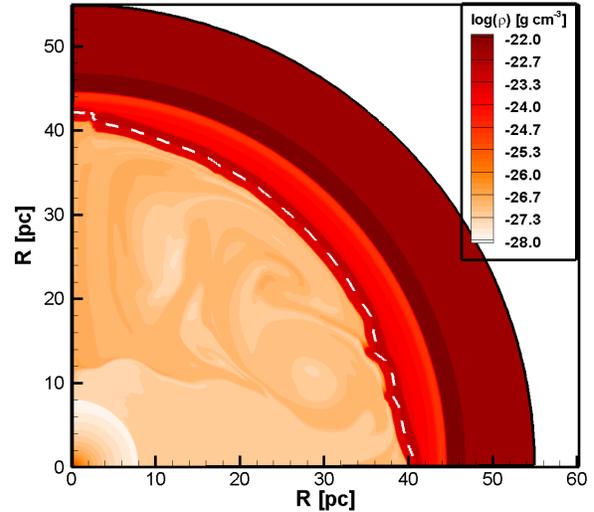}}
      \caption{Density distribution of the circumstellar medium $1.47\times10^3$yr 
  before the star explodes as a supernova
      (cf., Figs. \ref{fig:2d1} to \ref{fig:2d3}).
      The effects of the first critical rotation phase have long since disappeared. 
      The number of ionizing photons has decreased, while the ram pressure of the wind is 
larger than before the first period of critical rotation. 
      Together, these factors cause the hot bubble of shocked wind material to expand at the 
expense of the surrounding \ion{H}{II} region.
      The effect of the second period of critical rotation cannot yet be seen, since this 
becomes visible only in the last few hundred years before the supernova (See Sect. 
\ref{sec-result2}).
      }
     \label{fig:2dend}
   \end{figure}

\section{Results}

\subsection{Core hydrogen burning}
\label{sec-result1}
For most of the core hydrogen burning, the star rotates at about $\Omega =0.6$ (Fig.~\ref{fig:mloss1}),
which does not produce significant asphericities. Moreover, even though the wind termination
shock may be slightly aspherical, the main sequence bubble is pressure driven and will thus
be very close to a spherical shape anyway. For this reason, we modeled the evolution of the CSM of our
20$\mso$ star in 1D, up to shortly before the first sharp rise in $\Omega$ (Fig.~\ref{fig:mloss1}).
  For this we use a spherically symmetric grid, with 2000 radial grid points. 
  The Interstellar medium (ISM) density is set to ${10^{-22.5}{\rm g/cm}^3}$.
  The result of this simulation is shown in Fig. \ref{fig:ms}. 
  This result closely resembles that of a non-rotating 25~$\mso$ star (van Marle et al. 
\cite{MLG04}). 
  Moving outward from the star, we first encounter the free-streaming wind, then the shocked 
wind material, then the \ion{H}{II} region of photo-ionized ISM, the 
shell of shocked ISM material and finally the unperturbed ISM. 
  As in van Marle et al. (\cite{MLG04}), the hot shocked wind material and photo-ionized ISM 
have reached pressure equilibrium. 
  This is not always the case, since a strong wind can sweep up the entire \ion{H}{II} region, 
as was shown in Freyer et al. (\cite{FHY03}) and van Marle et al. (\cite{MLG06}).
  The star reaches critical rotation for the first time after about 1.2868$\times 10^7$~years 
(Fig.~\ref{fig:mloss1}).
  Therefore we start out 2D simulation about 30\,000 years before this time.

\subsection{Core hydrogen exhaustion through helium burning}
To switch from 1D to 2D, we have mapped the result of the 1D simulation onto a 2D grid,
which has 1000 radial and 200 angular grid points for an angular size of $\pi/2$ (Fig.~\ref{fig:2d1}).

As the star reaches critical rotation, the mass loss rate increases by nearly three orders of magnitude (from 
$10^{-6}$ to 
$10^{-3}$~$\msoy$), while the average wind
velocity decreases by about $600~\kms$.  This effect is largest for near-equatorial latitudes.  
Since the ram pressure of the wind varies considerably with latitude, the wind termination shock
becomes very aspherical.
  Due to the high mass loss rate, a new shell is formed at the wind termination shock. 
  Inside the wind termination shock, the free-streaming wind becomes strongly anisotropic, 
with most of the mass flow concentrated close to the equatorial plane (Fig.~\ref{fig:2d1}). 
since the 

  The period of critical rotation only lasts for about 10\,000~years. 
  Afterwards, the mass loss rate decreases and the wind becomes almost spherically symmetric 
once again. 
  The Wolf-Rayet wind sweeps up the disk, forming another shell, which moves outward (Fig.~\ref{fig:2d2}).
  As a result, the aspherical shell at the original wind termination shock is broken up and 
the remnants of both shells are pushed out into the hot bubble, where they will eventually 
dissipate. 
  The point where the wind breaks through the shell depends on the exact wind parameters 
during and after the period of critical rotation. 
  Effectively, the disk around the equator deflects part of the wind upward, which increases 
the ram pressure at higher latitude. 
  At the same time, the shell is thinner at higher latitude, since the mass loss rate was lower in 
that direction during the period of critical rotation.
  By the time the star explodes as a supernova ($t~\sim~1.331\times10^7$~yr), the remnants of 
the shell have completely disappeared (Fig. \ref{fig:2dend}). 

  In the phase between the two periods of critical rotation (the core helium burning phase of the 
star), the corrected number of ionizing photons is smaller than during the main sequence phase, while 
the ram pressure of the wind increases. 
  Due to the lower number of photons, the Str{\a"o}mgren radius recedes, leaving the original 
main sequence shell ($R~\simeq~45$~pc) unsupported. 
  This shell starts to dissipate, while a new shell is formed at the transition between 
shocked wind material and the former \ion{H}{II} region. 
  This shell, driven by the thermal pressure of the shocked wind material, starts to expand 
into the former \ion{H}{II} region. 
  This expansion goes quite fast, since the density in the former \ion{H}{II} region is lower 
than in the unperturbed ISM.
  A similar effect is found for the CSM around non-rotating massive stars during the 
Wolf-Rayet phase (van Marle et al. \cite{MLG05b}). 

  If the left hand image in Fig. \ref{fig:2d2} were to be 
rotated around the equator, its shape resembles the ring structure observed in \object{SN 
1987A}. 
  Since it is a temporary feature which quickly disappears, it may indicate that the 
rather peculiar shape of \object{SN 1987A} is the result of a period of rapid rotation of the progenitor star some 
time in the recent past. 

\subsection{Beyond core helium burning}
\label{sec-result2}
The stellar wind evolution during the final $\simeq 2000\,$yr of the star's life as shown 
in Figs. \ref{fig:mloss1} and \ref{fig:mloss2} produces dramatic changes in the
CSM in the inner few parsecs around the central star.  
  In Fig. \ref{fig:2dend}, the final frame of our first 2D-hydro run, some non-spherical features 
can be seen close to the star, while the rest of the CSM remains spherically symmetric.
The effects of the first phase of near-critical rotation have disappeared.
The further evolution is followed by a second 2D simulation with higher time and spacial resolution,
and restricted to the inner 5\, pc of the CSM. Due to the short remaining time to core collapse,
the outer parts of the bubble will remain largely unchanged. 
   
\begin{figure*}
   \centering
   \mbox{\subfigure{\epsfig{figure=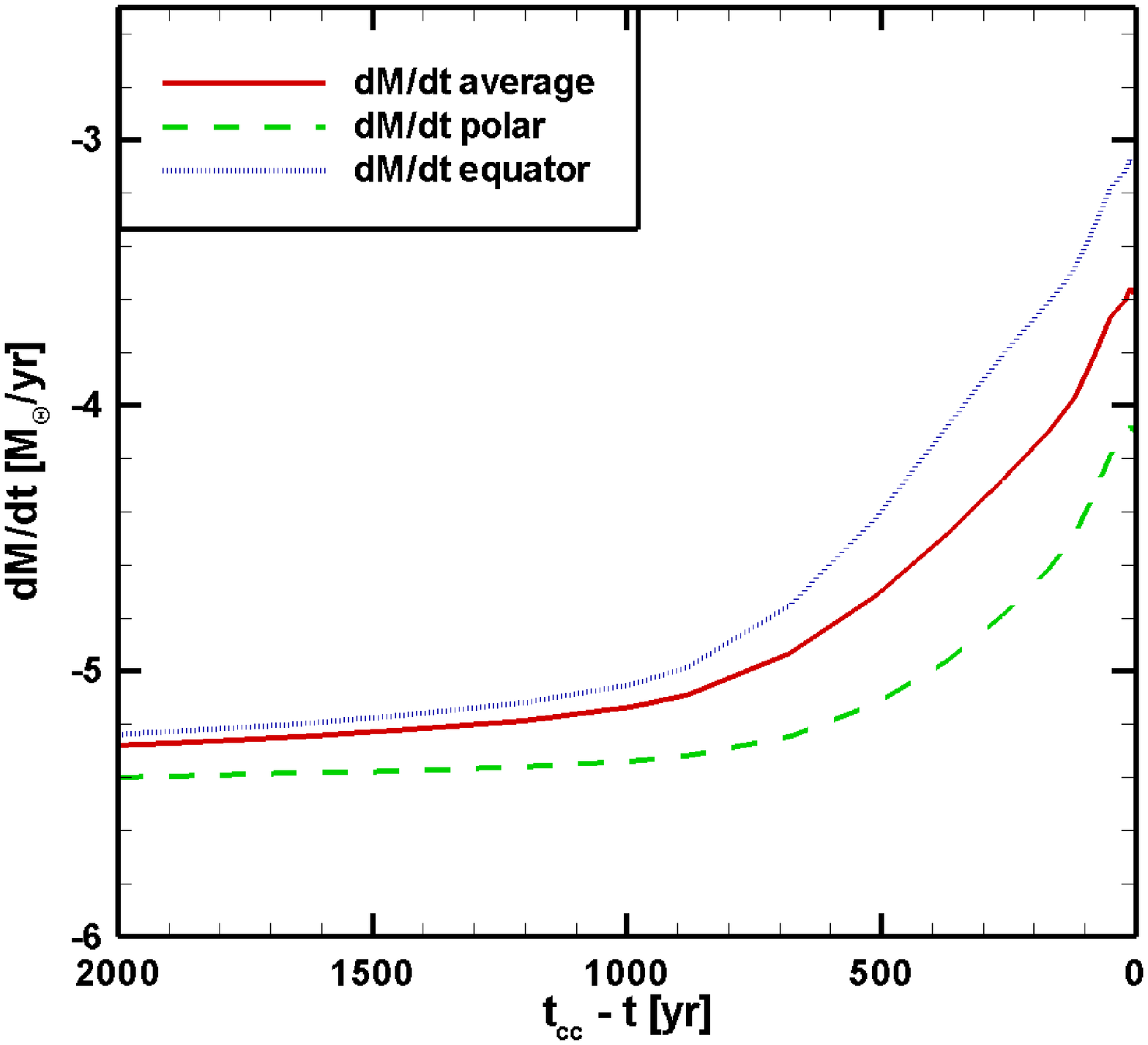,angle=-90,width=0.50\textwidth}}\quad
         \subfigure{\epsfig{figure=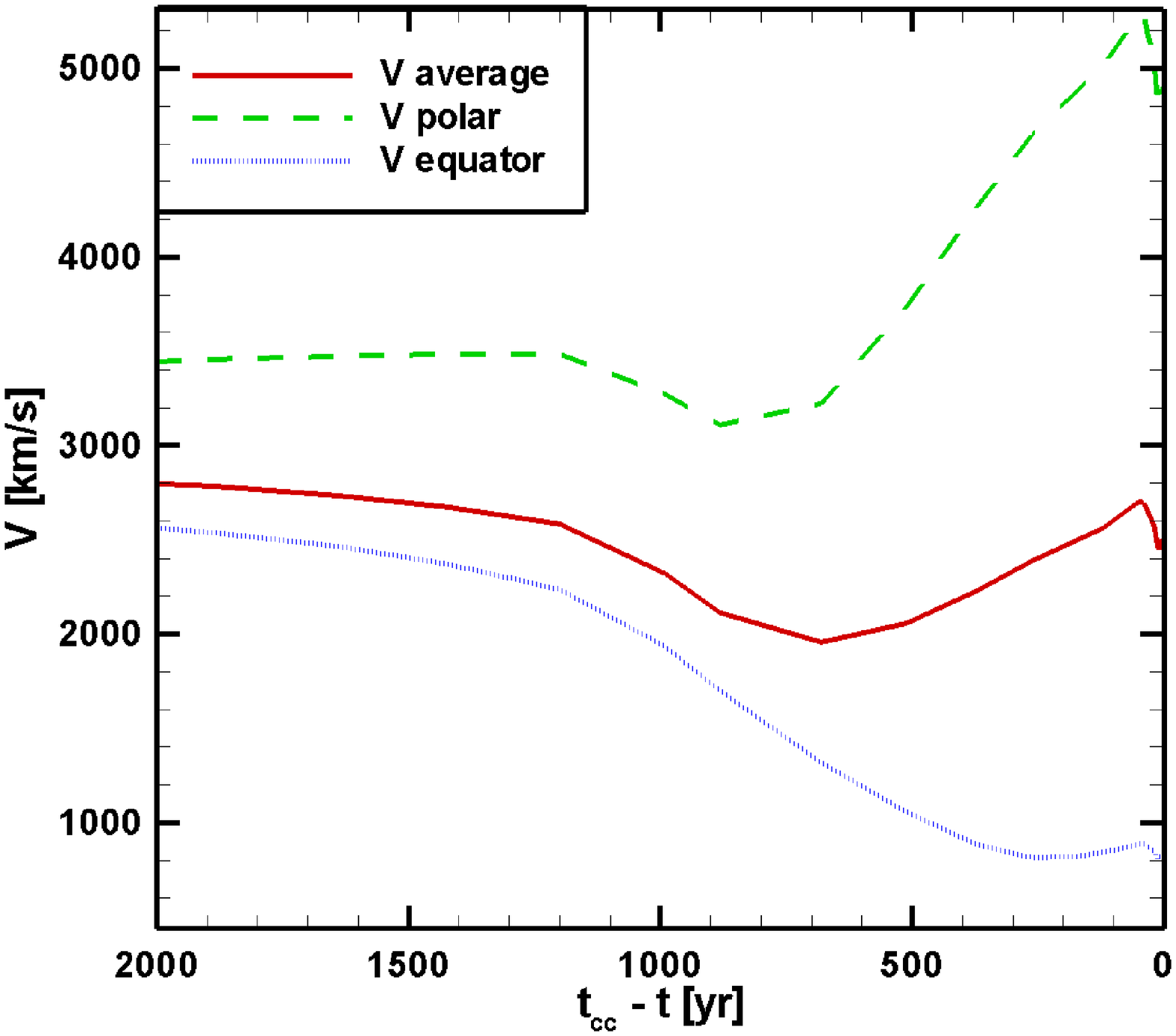,angle=-90,width=0.50\textwidth}}
	 }
	\caption{Mass loss rate and terminal wind velocity as function of time, for our 20 $\mso$
star, during the last 2\,000 years before core-collapse.
	The star approaches critical rotation and the mass loss rate increases considerably.
	The wind velocity first decreases as the period of rapid 
rotation starts, than it increases again as the star starts to shrink.
	The continuous lines show the angle averaged values, similar to Fig.~\ref{fig:mloss1};  
	The dotted lines show the mass loss rate and velocity along the polar axis, while the dashes lines give 
the values for the equatorial plane. 
	}
   \label{fig:final_1}
   \end{figure*}

\begin{figure}
   \centering
   \resizebox{\hsize}{!}{\includegraphics[angle=-90]{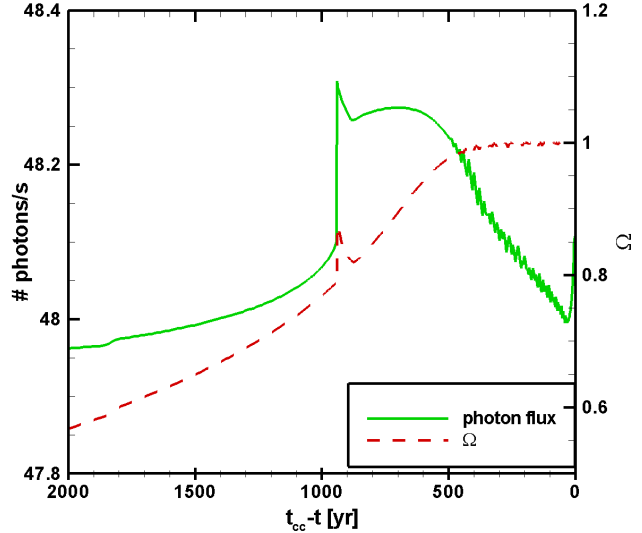}}
	\caption{Number of ionizing photons, and $\Omega$ ($=\vrot/\vcrit$), as a function of time,
during the same phase as shown in Fig. \ref{fig:final_1}. We use the average photon flux for all angles.
	}
   \label{fig:final_2}
   \end{figure}

  Just before the supernova explosion, the star reaches critical rotation for the second time 
(Fig. \ref{fig:mloss1}).
  This period lasts only about 1000\, yr, so no shell is formed at the wind termination shock, and 
the effects of the increased rotational velocity are not felt at large radius (see also Sects. 
\ref{sec-num} and \ref{sec-result1}).
  However, close to the star, the wind is aspherical once again.
  In Figs. \ref{fig:final_1} and \ref{fig:final_2} we show the mass loss rate, $\Omega$, the high 
energy photon count, and the wind velocity during the last stage of the evolution of the star.

\begin{figure*}
   \centering
   \mbox{\subfigure{\epsfig{figure=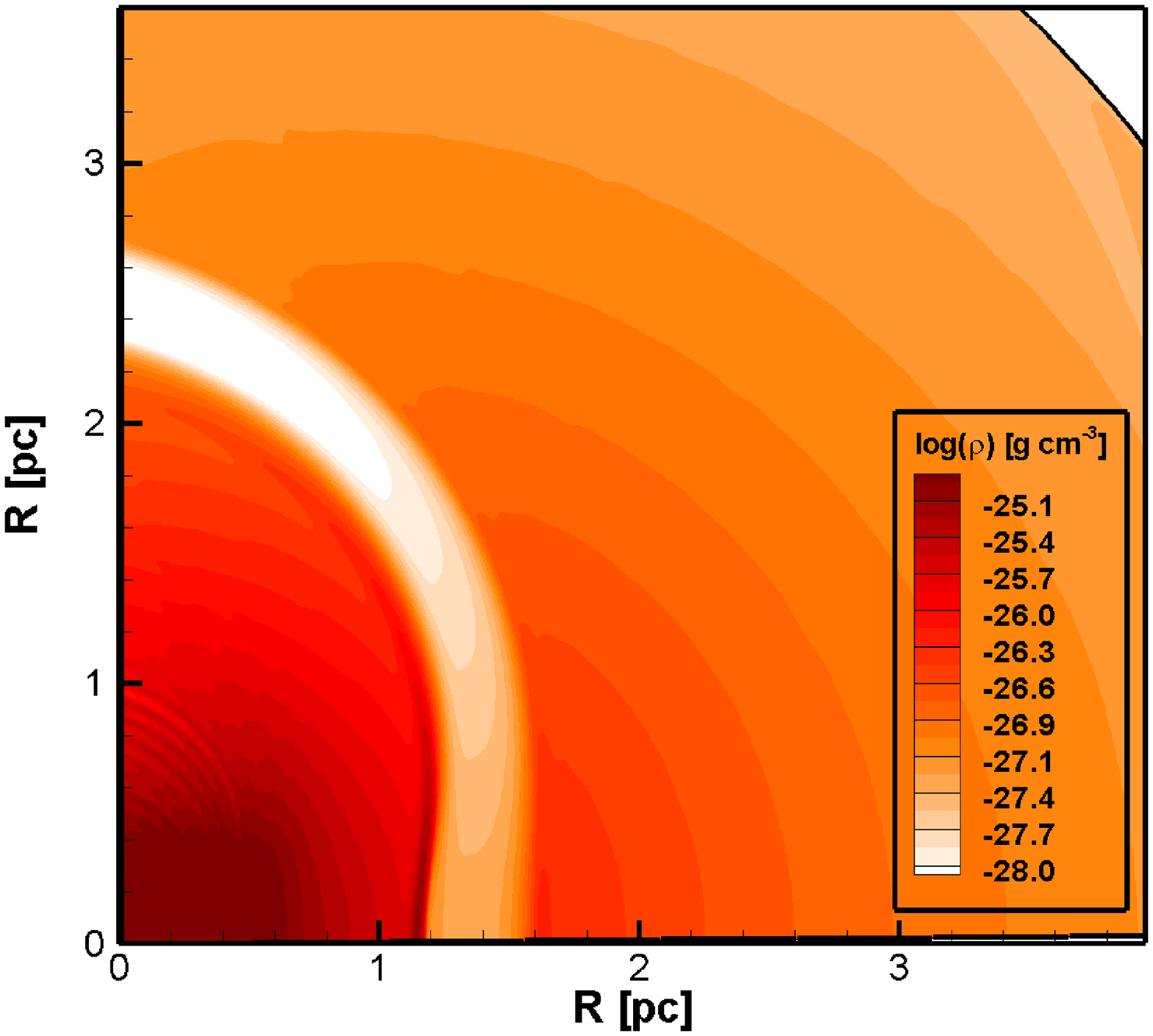,angle=-90,width=0.50\textwidth}}\quad
         \subfigure{\epsfig{figure=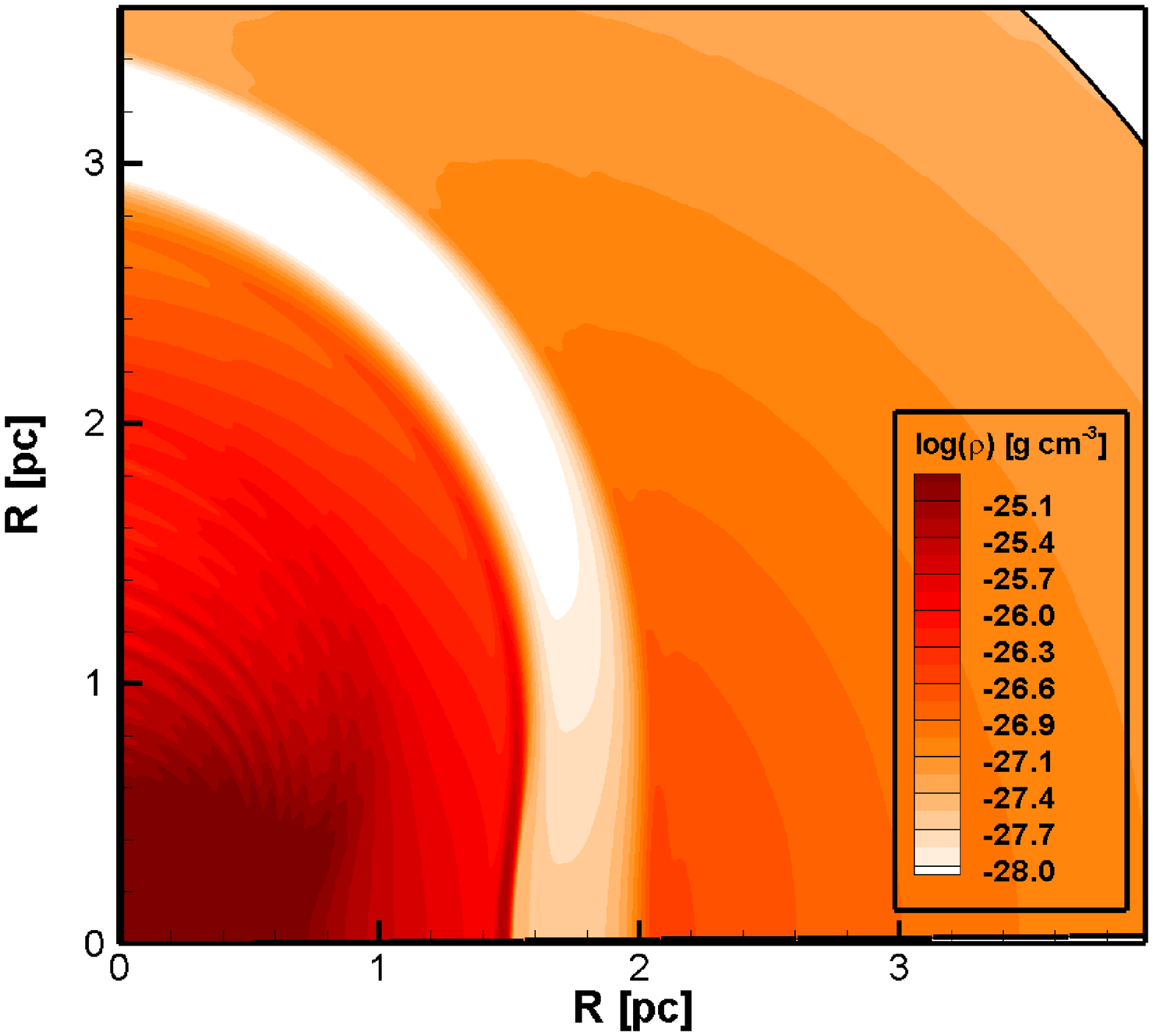,angle=-90,width=0.50\textwidth}}
          }
      \caption{Similar to Figs.~\ref{fig:2d1} and \ref{fig:2d3}. 
     The final evolution of the circumstellar medium close to the star. 
     These frames show the density of the circumstellar medium 21 yr and 1 yr pre-core collapse. 
    The star is rotating at critical velocity, creating a disk of matter around the equator.
     The wind velocity has decreased, which creates a rarefaction wave that moves away from 
the star ($R~\sim~2.3..2.7$~pc on the left, $R~\sim~3.0..3.5$~pc on the right).
     On the inner side of the rarefaction wave, a shell is forming due to the increased mass 
loss rate.
     At larger radii the disk starts to expand due to its internal thermal pressure. 
     Small changes in the escape velocity of the star cause a series of small waves in the 
wind material.
     The Str{\a"o}mgren Radius of the star lies outside the area of this simulation (Fig. 
\ref{fig:2dend}).
     }
     \label{fig:zoomin}
   \end{figure*}

The evolution of the CSM close to the star during the last 500~years is shown in Fig.~\ref{fig:zoomin}.
  Effectively, it repeats the evolution of the first stage of critical rotation. 
  At first, the wind velocity decreases (Fig. \ref{fig:final_1}).
  This creates a rarefaction wave, which moves outward in the free-streaming wind. 
  Then the wind becomes strongly aspherical and the mass loss rate increases (Figs.~\ref{fig:final_1} and 
\ref{fig:final_2}). 
  At the same time, the wind velocity increases again.
  A shell starts to form at the inner edge of this rarefaction wave, due to the increased mass 
loss rate and the fact that the new, faster wind starts to sweep up the material in the 
rarefaction wave. 
  However, before the shell has an opportunity to fully form, the evolution of the star is 
over and it explodes as a supernova, leaving Fig.~\ref{fig:zoomin} as the final morphology of the 
CSM close to the star, into which the GRB and supernova will expand. 

The $1/R^2$ density distribution of the free streaming wind is compromised, not only by the 
rarefaction wave, but also by small changes in mass loss rate and wind velocity during the final years of the 
stellar evolution.

\section{Velocity-resolved column depth}
\label{sec-spectrum}
  In the afterglow of \object{GRB~021004} a number of blue-shifted absorption lines were 
observed, indicating the presence of matter moving away from the progenitor star 
at discrete velocities. 
  van Marle et al. (\cite{MLG05b}) showed that these absorption features
might be explained by the hydrodynamic interactions in the circumstellar medium, as proposed by 
Schaefer et al. (\cite{Setal03}), Mirabal et al. (\cite{Metal03}), Fiore et al. 
(\cite{Fietal05}) and Starling et al. (\cite{Setal05}). 
  Since the rapidly rotating star has a different evolution from that of a non-rotating star, 
we expect to see a different absorption pattern here.

  In order to calculate the blue-shifted absorption pattern we use the same method as in van 
Marle et al. (\cite{MLG05b, MLG07}).
  This gives us the column density per 1~km/s interval as a function of radial velocity, which 
corresponds to the absorption as a function of blue-shift. 
  We correct for the Maxwell-Boltzmann distribution of particle velocities according to the
temperature of the gas.
  It is clear from Fig. \ref{fig:2dend} that the first phase of critical rotation will have no 
observable consequences at the time of the star's death. 
  Therefore, we concentrate on the CSM interactions due to the {\it second} period of 
critical rotation, described in Sect. \ref{sec-result2}. 

While van Marle et al. ({\cite{MLG05b, MLG07}) produced the absorption pattern as a function of time averaged over 
all angles, this is not useful here due to the strong asphericities in the CSM. 
For the GRB context, we are primarily interested in the potential absorption pattern in a GRB 
afterglow, which means that we focus on the material that lies along the 
polar axis of the star. 
  Therefore, we average the absorption pattern found along the first 20 radial grid lines, 
which corresponds to a cone with an opening angle of 18$^{\circ}$ around the polar axis (Fig.~\ref{fig:diagram1}).

  Figure~\ref{fig:coldens} shows the column density as a function of radial velocity for 
the final 1500\, yr of the evolution of the star. 
  In order to consider also the material beyond a radius of 5\, pc, we have combined the final outputs of 
both 2D simulations as follows.
  For the inner 5~pc, we use the final output shown in Fig.~\ref{fig:zoomin}. 
  For the rest of the circumstellar bubble, we use the output shown in Fig.~\ref{fig:2dend}.

  A strong component at zero velocity is always visible. 
  This originates from the ISM, the swept up shell of ISM material and the hot bubble, 
  where the material in the hot bubble has such a low density that it hardly contributes 
at all.
  Moreover, it has a high temperature, which smears its contribution out over an extremely 
large velocity interval. 
  The second feature occurs at high velocity and changes rapidly over time.
  
  At the start of Fig. \ref{fig:coldens} we only see a single high velocity component, since 
the stellar wind has a nearly constant velocity for some time. 
  The velocity then evolves according to Fig.~\ref{fig:final_2}. 
  First it drops almost instantaneously at $t~\sim~1.330595\times10^7$~yr. 
  After this, the velocity increases again. 
  This is especially true at the pole, since the rotational velocity increases also, and the 
stellar wind velocity for our rotating star is highest at the pole (Eq.~\ref{eq:vel}).
  Therefore, at the time of core collapse, we see three high velocity components at distinct velocities: 
  \begin{enumerate}
  \item The original wind velocity, which still exists in the free-streaming wind further away 
from the star. 
  \item The lowest velocity reached by the wind during this episode, which exists in the 
rarefaction wave that is moving outward.
  \item The latest wind velocity, which exists in the wind closest to the star. 
  \end{enumerate}
  The first two components become weaker over time, since the material with the original wind 
velocity enters the hot bubble where it is diluted and the rarefaction wave is being swept up by 
the faster wind that succeeds it. 
  
According to this model,
an observer might see a double or triple absorption feature at high velocity. 
  A double absorption feature was in fact observed for \object{GRB~021004} (Fiore et al. 
\cite{Fietal05}, Starling et al. \cite{Setal05}). 
  Absorption features at intermediate velocities (100..600~km/s), which were also observed in 
\object{GRB~021004}, are not produced in our model. 
For those to occur, the wind velocity in the phase of near-critical rotation
would have to drop to well below 1000~km/s, and an Homunculus-type nebula might have been
produced (Langer et al. \cite{LGM99}). 
Whether such low wind velocities can be produced in the final stage of
chemically homogeneously evolving stars of different mass and/or metallicity will
be investigated in the near future.
The ionization state of the circumstellar medium is not taken into account in our calculations. 
If the gas were ionized to a high degree, spectral features like \ion{C}{IV} and Si{IV} would not be visible at 
all 
(Prochaska et al. \cite{PCB06}, Chen et al. \cite{Cetal07}).
{\bf Therefore the spectral lines observed in \object{GRB~021004} would be produced far away from the progenitor 
star, indicating an uncommonly large free-streaming wind region, or be unrelated to the star. Alternative 
explanations for this contradiction may be the presence of dust in the stellar wind, which influences the 
ionization state of the gas, or a structured GRB beam (Starling et al. \cite{Setal05}), which would produce an 
afterglow over a much broader front than the initial GRB}

   \begin{figure}
   \centering
   \resizebox{\hsize}{!}{\includegraphics[width=0.95\columnwidth]{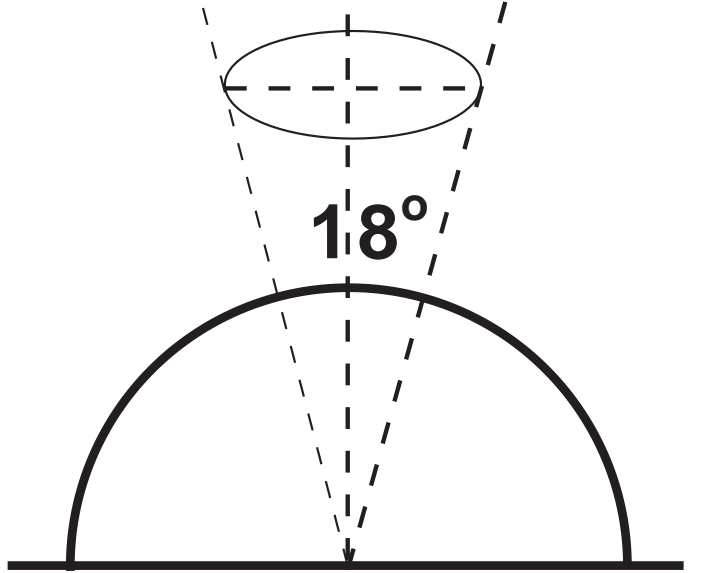}}
      \caption{To produce Fig. \ref{fig:coldens}, we average the the ``absorption spectra'' 
      over each radial grid line in a 18$^{\circ}$ 
cone around the polar axis.
      }
     \label{fig:diagram1}
   \end{figure}

   \begin{figure}
   \centering
   \resizebox{\hsize}{!}{\includegraphics[width=\columnwidth,angle=-90]{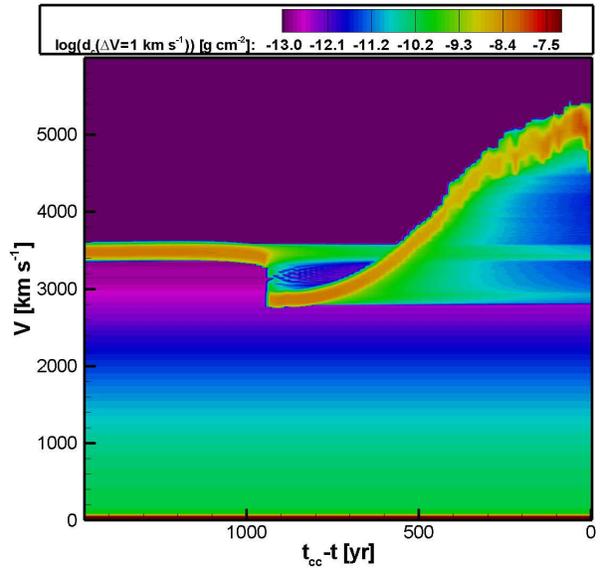}}
      \caption{Column density per radial velocity interval of 1~km/s, as a function of radial velocity and 
time, for the final 1500\, yr of the evolution of the star. The plot ends at the time of core collapse.
The horizontal axis shows the time prior to core collapse ($t_{cc}$).
      The changes in wind velocity during the last thousand years of the evolution produce a 
complex spectrum of high velocity absorption components.      
      }
     \label{fig:coldens}
   \end{figure}
     
\section{The circumstellar density profile}
\label{sec-dprof}
  The density of the circumstellar medium directly influences the shape of the GRB afterglow 
light curve. 
  Numerical and analytical models show that many GRBs seem to occur in a medium that does not 
conform to the $1/R^2$ density profile that one would expect for a free streaming wind of a massive star 
(Chevalier \& Li \cite{CL00}; Panaitescu \& Kumar \cite{PK01}, \cite{PK02} and Chevalier et 
al. \cite{CLF04}). 
With $\alpha=d\log \rho / d\log r$, small values of $\alpha$ can be 
explained by assuming that the GRB jet runs into the shocked wind, 
rather than into the freely expanding wind where the density is close to constant ($\alpha=0$).
However, this requires the wind termination shock to be 
very close to the star (Wijers~\cite{W01}; van Marle et al.~\cite{MLAG06}, 
\cite{MLAG07}). 

  The location of the wind termination shock depends on the balance between the ram pressure 
of the wind and the thermal pressure of the hot wind bubble that contains the shocked wind. 
The bubble pressure depends primarily on the mechanical luminosity of the stellar wind 
during all the preceding stages of the stellar evolution. 
  Therefore in order to bring the wind termination shock close to the star, the ram pressure 
of the wind during the final stages must be relative low, compared to the ram pressure of the 
wind during earlier phases. 
  For a normal Wolf-Rayet type star, this is difficult to achieve, since Wolf-Rayet star winds 
tend to have both high velocity and a high mass loss rate. However, the chemically 
quasi-homogeneous model described in this paper has a very low metallicity and a 
completely different evolution. 
  As a result of the lower metallicity, the Wolf-Rayet wind is not as strong as for solar metallicity 
stars.
  The thermal pressure in the shocked wind bubble is not very different form the high metallicity case, 
since it depends on the 
mechanical luminosity of the wind to the power 0.4 (Weaver et al. \cite{WCMSM77}).
  Therefore, the radius where those two pressures are in balance will be closer to the central star 
for low metallicity stars.
  A second factor is the lack of a red supergiant phase. 
Initially, a red supergiant wind constrains the wind termination shock of the Wolf-Rayet wind, which has to plow 
through the dense material of its predecessor (Garc{\'i}a-Segura et al. \cite{GLM96}, van Marle et al. 
\cite{MLG05b}). 
However, once the Wolf-Rayet wind driven shell reaches the hot bubble created by the main sequence wind, a new 
equilibrium is reached between the thermal pressure in this bubble and the Wolf-Rayet wind ram pressure. 
Since the red supergiant wind has a low mechanical luminosity (due to low wind velocity) it contributes very 
little to the thermal pressure of the hot bubble while still giving it time to cool and expand, causing the 
thermal pressure to drop quickly, which places the wind termination shock further away from the star. 
This means that, unless the GRB occurs before the Wolf-Rayet wind reaches the hot main sequence bubble, which 
typically happens within about 25\,000 years (Garc{\'i}a-Segura et al. \cite{GLM96}, van Marle et al. 
\cite{MLG05b}), a red supergiant phase will make it more difficult to create a constant density medium close to 
the star. 
Since chemically near-homogeneous stars do not have a red supergiant phase, their wind termination shocks will 
tend to be closer.
  In Figs.~\ref{fig:2d3} and \ref{fig:2dend} we can see that the wind termination shock lies 
approx. 5-10 pc from the progenitor star. 
  This is about 30-50\% of the radius of the wind termination shock for a 'normal' Wolf-Ratet 
star in a similar environment (van Marle et al. \cite{MLAG06}).
  In itself, this is not enough to explain the presence of a constant density medium in the GRB 
afterglow but it can be combined with external factors, such as stellar motion and high 
density interstellar medium to bring the wind termination shock close enough to the star (van 
Marle et al. \cite{MLAG06}).
  
   \begin{figure}
   \centering
   \resizebox{\hsize}{!}{\includegraphics{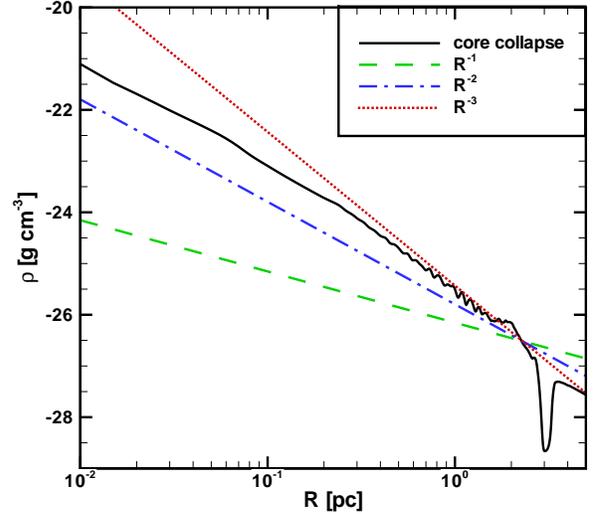}}
      \caption{Density profile average over a cone with opening angle 18 degrees around the polar 
axis, at the time of core collapse. The density profile does not conform to a $R^{-2}$ 
distribution, as expected from a constant wind. Density profiles with the density varying as $R^{-1}$, 
$R^{-2}$, and $R^{-3}$ are shown for comparison.
      This plot also shows the rarefaction wave caused by the sudden drop in wind velocity just prior 
to the period of critical rotation (See Fig.~\ref{fig:final_2}).
      }
     \label{fig:dprof}
   \end{figure}

As evident from Fig.~\ref{fig:2dend}, the final density profile in the inner few parsec of the CSM
of our 20 $\mso$ star is not well described by an $\alpha=-2$ law. 
  In Fig.~\ref{fig:dprof} we show the final density profile along the polar axis produced by our
simulations.
  This density profile is obtained as the average over a cone with an opening angle of eighteen degrees around the
polar axis, equal to the angle used in Figs.~\ref{fig:diagram1} and \ref{fig:coldens}.
While the density follows $\alpha=-2$ up to about $R=0.2\,$pc, the density slope steepens
considerably, up to $\alpha=-3$ beyond that radius, up to the steep density drop 
at $R\simeq 3\,$pc caused by the rarefaction wave (see Fig.~\ref{fig:final_2}). 

While we can not generalize from the single case that we investigate here, our results show that
strong deviations from a density slope of $\alpha=-2$ may be expected even in the regions very 
close to the GRB progenitor, well inside the classical wind termination shock. The reason is
the short evolutionary time scale of the star itself, after core helium exhaustion, and the fact that
near-critical rotation is reached in this phase, which greatly amplifies the sensitivity of the
wind properties to the changes in the global stellar properties. While in our example case we found
a decrease of the density slope in the inner few parsec as consequence of an increased wind speed in the
last $\sim 1000\,$yr, cases where the rotation of the stellar surface drops well below $\Omega\simeq 1$
during the last few thousand years of the star's life are also found in the GRB progenitor evolution
grid of Yoon et al. (\cite{YLN06}), where $\alpha$ values closer to zero may be expected.

\section{Conclusions}
\label{sec-concl}
The time evolution of the mass loss rate and wind velocity of 
a rapidly rotating, chemically homogeneous massive star differs 
radically from the evolution of a non-rotating star.  
  This difference is reflected in the evolution of the CSM. 
  As was shown in Sects. \ref{sec-result1} and \ref{sec-result2},
our rapidly rotating 20~$\mso$ star produces a large
bubble, which resembles that of non-rotating stars. 
  However, during the two periods of near-critical rotation, large-scale deviations from
spherical symmetry are produced. 
  While the outer edge of the bubble remains spherical, closer to the star the asphericity 
of the wind creates unique morphologies. 
  As the second phase of near-critical rotation occurs immediately before the supernova explosion, 
the supernova itself, and a possible GRB, will expand into such an aspherical medium.
This will clearly influence the supernova remnant evolution, and the GRB afterglow light curve.
 
For a GRB, the effect of a non-spherical CSM can be felt in two ways.
  First, the relativistic jet will not expand in a normal, free streaming wind, but in a 
medium that has a far more complex structure (Sect.~\ref{sec-result2}). 
In particular, the density slope in the innermost few parsec, well inside the wind termination shock,
may differ significantly from $R^{-2}$, and may also show significant changes as function of radius.
bf As the changes of the stellar wind during the last few 10000 yr
depend on the internal structure of the star at this stage, it requires
the consideration of a large grid of stellar evolution models to
find out what the typical final density profiles would be, and
how large their expected variation is

  Second, the medium in front of the jet does not have a constant velocity. 
  Assuming that not all of this material is ionized by the GRB, it may be observable 
in the form of blue-shifted absorption lines (Sect. \ref{sec-spectrum}).
  Our investigation shows that the wind-wind interactions during the final stages of the 
stellar evolution produce multiple absorption lines at high velocity. 
  Such lines were observed in the afterglow of \object{GRB~021004}.
While it is debated as to whether these absorption lines are in fact circumstellar in origin 
(Prochaska et al. \cite{PCB06}, Chen et al. \cite{Cetal07}), the possibility of producing them 
in this way is shown here for the first time. Furthermore, such lines might also be observed in 
early-time spectra of nearby Type~Ib/c supernovae at low metallicity.

\begin{acknowledgements} 
      We would like to thank Wim Rijks at SARA Computing and Networking Services 
(http://www.sara.nl/) for parallelizing the numerical code.\\    
      This work was sponsored by the Stichting Nationale Computerfaciliteiten (National 
Computing Facilities Foundation, NCF), with financial support from the Nederlandse Organisatie 
voor Wetenschappelijk Onderzoek (Netherlands Organization for Scientific research, NWO).
      This research was done as part of the AstroHydro3D project:\\
      (http://www.strw.leidenuniv.nl/AstroHydro3D/)\\
      A.J.v.M acknowledges support from NSF grant AST-0507581 \\
      S.-C.Y. is supported by the VENI grant (639.041.406)
of the Netherlands Organization for Scientific Research (NWO).
\end{acknowledgements}

\end{document}